# Interface Selection for Power Management in UMTS/WLAN Overlaying Network

*Mostafa Zaman Chowdhury[†], **Yeong Min Jang[†], Choong Sub Ji[†], Sunwoong Choi[†],
Hongseok Jeon[††], Junghoon Jee[††], and Changmin Park[††]
[†]Kookmin University, Korea
[††]Electronic and Telecommunications Research Institute (ETRI), Korea
*mzceee@yahoo.com, **yjang@kookmin.ac.kr

*Abstract* — The multiple choices of access networks offer different opportunities and overcome the limitations of other technologies. Optimal selection of interface is a big challenge for multiple interfaces supported mobile terminals to make a seamless handover and to optimize the power consumption. Seamless handover, resource management, and CAC to support QoS and multiple interface management to reduce power consumption in mobile terminal are the most important issues for the UMTS/WLAN overlaying network. The access of both interfaces simultaneously can reduce the handover latency and data loss in heterogeneous handover. The MN may maintain one interface connection while other interface can be used for handover process. But the access of both interfaces increases the consumption of power in MN. In this paper we present an efficient interface selection scheme including interface selection algorithms, interface selection mechanism and CAC considering battery power consumption for overlaying networks. MN's battery power level and provision of QoS/QoE in the target interface are considered as important parameters for our interface selection algorithm. The MIH is also introduced for interface selection.

## 1. Introduction

The use of wireless communication has been increased tremendously in the recent years and it will continue in the future. Due to these huge demands, varieties of user types and varieties of user's requirement, different wireless technologies have been developed. These technologies vary widely in terms of bandwidths, QoS provisioning, security mechanisms, price, coverage area and etc. The complementary characteristics of WLANs and Universal Mobile Telecommunications System (UMTS) based cellular networks make them attractive for integration. This integration offers the best of both technologies. Thus a mobile node (MN) with multiple wireless interfaces has become increasingly popular in recent years [6]. In heterogeneous overlay network, the MN can select one interface that is best or suitable in terms of price, Quality of Service (QoS), Quality of Experience (QoE), throughput or other parameters as required. During connection, due to changes in the availability or characteristics of an access network may result in a situation where already established connections should be moved from one interface to another.

But for a MN, especially a battery-operated device with multiple wireless interfaces, power consumption is one of the critical problems [6]. Figure 1 shows that the uses of multiple interfaces consume more power than that of use just single interface. Also the access of different network interface causes different level of power consumption.

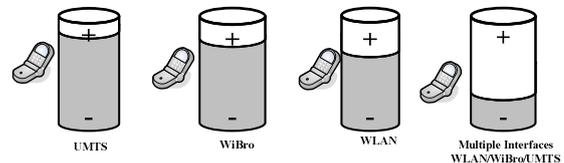

Figure 1.  Battery power consumption for the use of different interfaces

Traditionally for horizontal handovers, only signal strength and available bandwidth are used as handover decision parameters. Also for traditional overlay network, the handover decision depends on several parameters like signal strength, available bandwidth, price of the link, security level, and coverage radius [4]. As power of the battery is very important issue for a MN with multiple interfaces, battery power level of a MN should be considered as important parameter for suitable interface selection. Thus power management issue can be added to IEEE 802.21 Media Independent Handover (MIH) [1], [3] for interface selection in overlay network. For our proposed interface selection algorithm, MN's battery level and provision of QoS/QoE in the target interface have been considered as important parameters for our interface selection.

The power consumption of a MN depends on received signal strength and type of access network. So, the proper design of call admission control (CAC) procedure is also essential to reduce the power consumption of the MN.

This paper is organized as follows. Section 2 provides the brief description about the power management issues in IEEE 802.21 MIH. Relative works as well as our proposed algorithm for interface selection are presented in Section 3. In Section 4, we propose interface selection steps, interface selection functional architecture and a CAC for efficient interface selection. The numerical results for the proposed algorithm are presented in Section 5. We give our conclusion in Section 6.

## 2. MIH and Power Management

The purpose of 802.21 is to facilitate the handover between different interfaces (such as 802.11, 802.16, 3GPP, 3GPP2,

and 802.15) and provide a handover management scheme in such a way that is independent from particular access network features [1], [3]. The MIH includes three different major functions. The event service initiates handover from the lower layer by means of low-layer trigger events, the command service initiates handover from the upper layer to control connectivity and the information service controls communications of basic static information [3].

The introduction of power management functionalities in MIH can improve the network selection performance. Figure 2 shows the power management functionalities in MIHF.

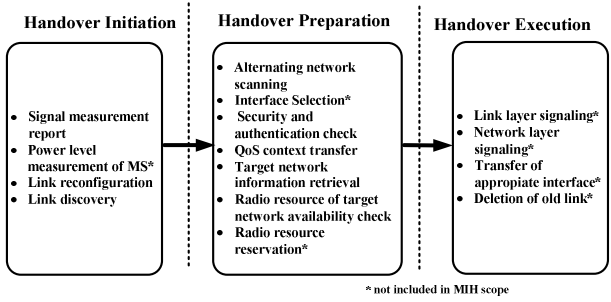

**Figure 2. Power management in MIHF**

## 3. Interface Selection Algorithms

### 3.1 Related Works and Background

There are several parameters those are used for handover decision. The interface selection procedure is a Multiple Attribute Decision Making (MADM) problem where alternative options are presented by multiple number of links (interfaces). Best handover decision depends on how the parameters are selected and how these parameters are used for interface selection algorithm. There are several works already done for this area. Different researchers [2]-[8] assume different parameters for their interface selection algorithm but no one assume the status of MN's battery or battery profile and provision of QoS/QoE in the target interface as their interface selection algorithm. We consider these two parameters for our proposed interface selection procedure.

The authors in [4] used six parameters for each interface; signal strength, bit rate, power consumption, price, coverage and security. They made weight vector or profile for the interface selection algorithm. The Simple Additive Weighting (SAW) and Weighted Product proposed in [4] for the measurement of each property.

$$S^t_{saw} = R^t_i \frac{\sum_{j=1}^{p} w_j * r'^t_{i,j}}{\sum_{j=1}^{p} W_j} \quad (1)$$

$$S^r_{wp} = R^t_i \prod_{j=1}^{p} (r'^t_{i,j})^{w_j} \quad (2)$$

Each property may have different intervals for its values.

$r'^t_{i,j}$ $\psi\psi$ represents the scaled $r^t_{i,j}$ $\psi\psi$ value.

Authors in [9] proposed cost function based model for interface selection algorithm. Signal strength (s), cost of using the network access technology (c) and client power consumed for the particular access technology (p) are used as input parameters for their algorithm. They use score function (SF) for interface selection decision.

$$SF_i = (w_s * f_{s,i}) + (w_p * f_{p,i}) + (w_c * f_{c,i}) \quad (3)$$

According to [8] the power consumption for a specific application in WLAN $C_w$ and power consumption in UMTS $C_u$ are given by

$$C_W = (P_{tw}C_{tw} + P_{rw}C_{rw} + P_{lw}C_{lw} + P_{sw}C_{sw})T \quad (4)$$

$$C_U = (P_{tu}C_{tu} + P_{ru}C_{ru} + P_{su}C_{su} + P_{pu}C_{pu})T \quad (5)$$

In equation (4) and (5) $C_{tw}$, $C_{rw}$, $C_{lw}$ and $C_{sw}$ represent the power consumption in transmit, receive, idle and sleep state respectively, while $P_{tw}$, $P_{rw}$, $P_{lw}$ and $P_{sw}$ are the probabilities of being in any of the respective communication state. $C_{tu}$, $C_{ru}$, $C_{su}$ and $C_{pu}$ represent the power consumption in transmit, receive, signaling and power-saving state respectively, while $P_{tu}$, $P_{ru}$, $P_{su}$ and $P_{pu}$ are the probabilities of being in any of the respective communication state. Hence power consumption also depends on different mode of operations.

### 3.2 Proposed Algorithm

In our proposed algorithm we divide all the interface selection parameters ($w_i$) into two groups. One group takes more priority than another group for interface selection decision. We have $N$ number of available interfaces. Suppose, battery power level and other $M$ parameters for interface selection and among them $q$ parameters have more priority than other remaining $(M-q)$ parameters, then the weight $(W)$ of the measurement is presented as

$$W = f_1 [w_1, w_2, ..., w_q] + f_2 [w_{q+1}, w_{q+2}, ..., w_M] \quad (6)$$

Suppose $S_m$ indicates the scaling factor of $m^{th}$ parameter and $L_p$ indicates the battery power level of the MN, then the weight of $p^{th}$ interface among $N$ interfaces is:

$$W_p = \frac{f(w_m, S_m)}{f(L_p)} \quad (7)$$

Current level or status of MN's battery condition should be considered as well for interface selection algorithm with other traditional parameters. For our proposed algorithm we consider battery power level and other seven parameters $(M=7)$; signal strength (1), throughput (2), power consumption (3), cost (4), cell coverage (5), QoS/QoE level (6), and security (7). Equation (8) measures the weight for each interface.

$$W_p = \frac{\sum_{m=1}^{7} w_m \cdot S_m}{\log 10(1 + L_p)} \quad (8)$$

where $1 \leq p \leq N$ and $\sum_{m=1}^{7} s_m = 1$

In Equation (8) $L_p = 1$, for battery power level greater than a threshold value; otherwise $L_p = K$. Here, K is the rank of the interface according to power consumption. For lowest power consuming interface, $K=1$ and highest power consuming interface $K=N$.

Equation (7) and (8) introduce battery power level condition in the interface selection. This parameter is only effective whenever the power level goes down a threshold battery power level. Lower than threshold level means, the battery power level is going to be worst condition and thus the MN should select an interface that consumes lower power. The impact of $L_p$ in the interface selection algorithm may be changed as operator desired.

## 4. Interface Selection Procedure

Figure 3 shows the steps for interface selection. The cross layer information for different interfaces are collected and then these information and some pre-defined policies for interface selection are checked using appropriate algorithm to make a best interface selection decision. The algorithm calculates total weight for each interface. According to result of the algorithm, all the available $N$ interfaces are ranked. For example best interface is ranked as 1 and the worst one is ranked as $N$. Thus MN will try to handover to best selected network. If resources are available in the best selected interface, then the MN handover to that interface otherwise it will try for the next ranked interface. This process will continue until $(N-1)$ ranked interface.

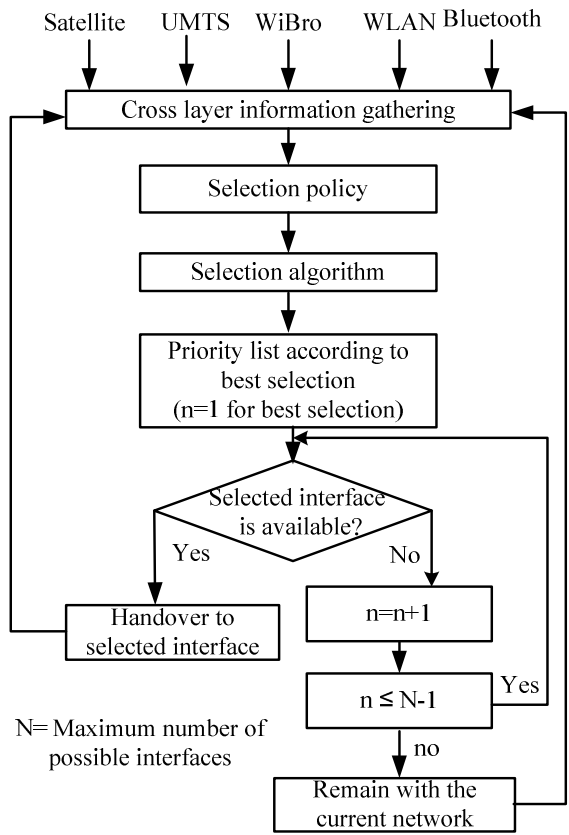

**Figure 3. Interface selection steps**

Figure 4 shows the interface selection functional architecture. The decision engine collect information from user interface, battery profile, policy engine, MIHF and link information engine. User interface provides information about the type of application, access technology, user's QoS/QoE requirement and etc. Battery profile provides the information about battery power level. Policy engine provides the pre-defined policies. The link information engine observes different layers condition and combine these information using cross layer optimization and then forward these information to decision engine. The decision engine selects a best interface and forwards the decision to handoff module. The decision engine also makes a rank for the available interfaces according to the total weight. The handover module executes the handover.

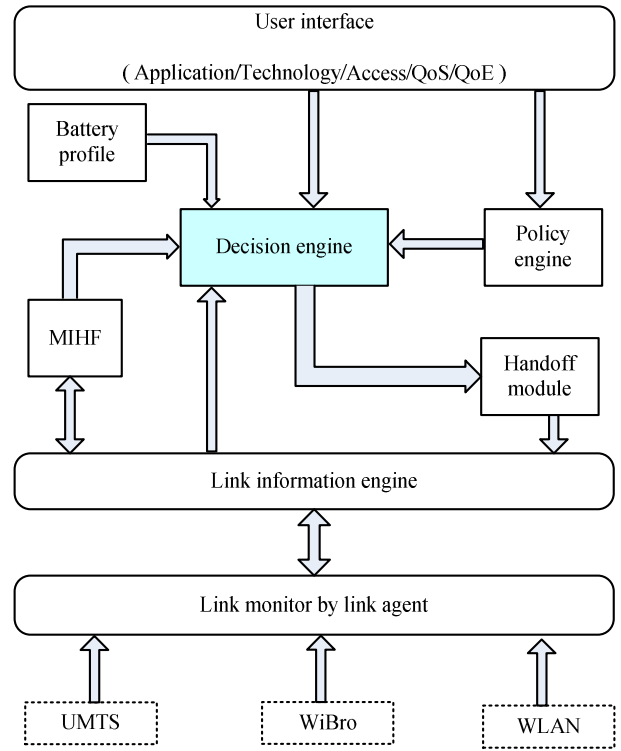

**Figure 4. Interface selection functional architecture**

Figure 5 shows an example of a CAC for power management based interface selection procedure in a UMTS/WLAN interworking. Handover from UMTS to WLAN and WLAN to UMTS both the cases are considered in this CAC. For UMTS networks, MN is located far away from the base station (BS) means lower signal level. Thus received signal level became low and MN needs more power consumption. Hence distance of a MN from BS is also considered in the CAC. If distance of the MN is larger than a threshold distance or battery level higher than a threshold level then MN will try for WLAN interface. Whenever a MN is connected to WLAN, it will remain with WLAN until battery level does not go down a threshold level and distance of the MN from the BS is not less then the threshold distance.

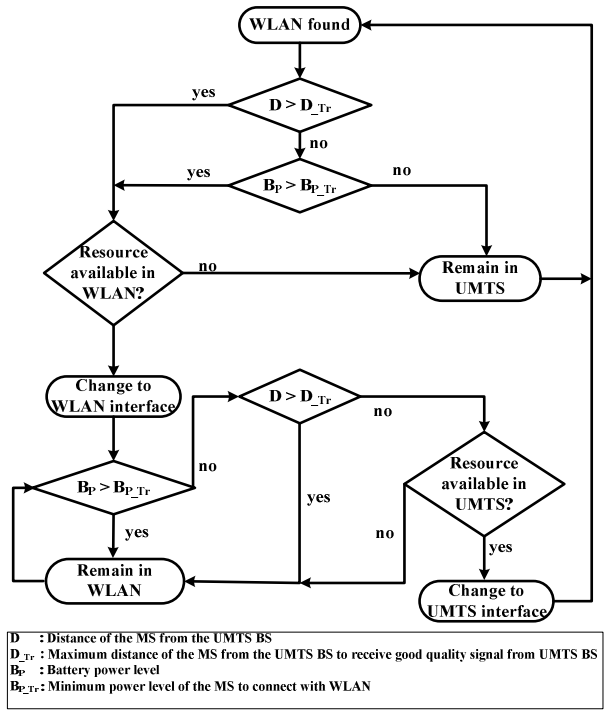

**Figure 5. CAC for interface selection**

## 5. Numerical Analysis

This section provides the results of our proposed interface selection algorithm. We consider Okumura-Hata model [10] for path loss calculation in our numerical analysis. Table 1 shows the basic assumption for Okumura-Hata model. We also assume that the distance of MN from every WLAN AP is 10m whatever the distance of MN from macrocell BS. Table 2 shows the basic assumption for the parameters those are considered for interface selection algorithm. The assumed scaling factor of each parameter is given here. In our numerical analysis we assume requested bandwidth is available in both the interfaces.

**Table 1. Parameters for path loss model**

| Access network | Parameter | Assumption |
|---|---|---|
| UMTS | BS transmit signal power | 1.5 KW |
| | Path loss model (Okumura-Hata model for macrocell) | $L_p = 69.55 + 26.16 \log f_c - 13.82 \log h_b - a(h_m) + [44.9 - 6.55 \log h_b] \log d$ |
| | Height of BS | 100m |
| | Height of MN | 2m |
| | Receiver sensitivity | -100 dB |
| WLAN | AP transmit signal power | 100 mW |
| | Path loss model (Okumura-Hata model for microcell) | $L_p = 135.41 + 12.49 \log f_c - 4.99 \log h_b + [46.84 - 2.34 \log h_b] \log d$ |
| | Height of AP | 2m |
| | Coverage area | 15 m |

**Table 2. Assumption for weight parameters**

| Parameter | Scaling factor | Weight ratio |
|---|---|---|
| Cost | 0.4 | UMTS(1): WLAN(10) |
| Throughput | 0.2 | UMTS(1): WLAN(10) |
| QoS/QoE | 0.09 | UMTS(1): WLAN(4) |
| Cell coverage | 0.05 | UMTS(100): WLAN(1) |
| Security level | 0.08 | UMTS(4): WLAN(1) |
| Signal strength | 0.08 | Depends on the distance of MN from UMTS BS |
| Power Consumption | 0.1 | Depends on the distance of MN from UMTS BS |

In our numerical analysis we calculate that the battery power consumption of a MN for UMTS interface is less than the WLAN interface if the distance of MN from UMTS BS is less than 920m. So for power saving mode, MN can select UMTS interface when the distance is less than 920m and for a distance more than 920m, the MN can select WLAN interface to connect the user for longer time.

As we assume that the distance of MN from every WLAN AP is 10m whatever the distance of MN from macrocell BS, the received signal strength from the WLAN AP can be considered as same value for every WLAN environment. Hence the weight of WLAN interface is almost constant for normal operating condition. But with the increase in distance between UMTS BS and MN, the received signal level decrease and also battery power consumption increase. So, the total weight of UMTS interface decreases with the increase of distance. Figure 6 shows the total weight of each interface whenever MN has sufficient battery power level. It shows that WLAN is better due to low cost and higher throughput.

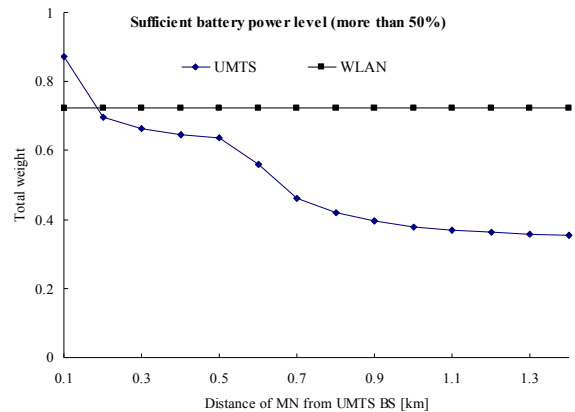

**Figure 6. Total weight of each interface whenever MN has sufficient battery power level**

Figure 7 shows the total weight of interface selection whenever battery power level is not sufficient. At this moment saving of power is more important. Thus MN can connect with the wireless link for longer time using power saving mode. Figure 7 also shows that MN should select UMTS interface when distance is less than 600m by considering all the parameters. For the distance more than 600m, the MN should select WLAN interface to reduce the power consumption if both the interfaces are available.

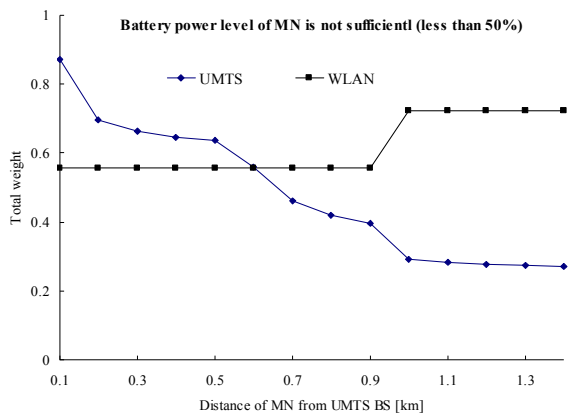

Figure 7. Total weight of each interface whenever MN has insufficient battery power level

## 6. Conclusion

Multiple choice of interface selection is a good opportunity to access multiple access networks with the suitable price and better QoS/QoE level as required. The main problems with the multi-mode operated MN are very high battery consumption and difficulties in the selection of best interface. In this paper we proposed some new functionality in MIH to save power. The current battery power level has been considered as the interface selection parameter for the interface selection algorithm. Thus for lower battery level environment, the MN will select an interface that consume less power. We also considered QoS and QoE level that can be provided by target network in the interface selection algorithm. The proposed functional architecture for interface selection and CAC can provide a best handover decision for overlaying network. The numerical results show that the proposed algorithm is capable to select appropriate interface in both the normal operating mode and power saving mode. The MN will support the seamless services for longer time by the proposed power saving mode operation.

## Acknowledgement

This research was supported by the MKE (Ministry of Knowledge and Economy), Korea, under the ITRC (Information Technology Research Center) support program supervised by the IITA (Institute of Information Technology Assessment) (IITA-2008-C1090-0801-0019). This research was also supported by Electronic and Telecommunications Research Institute (ETRI), Korea.